\PassOptionsToPackage{dvipsnames}{xcolor}
\documentclass[preprint]{aastex62-no-date}
\pdfoutput=1
\usepackage{amsmath}
\usepackage{capt-of}   % don't use the package caption -- it causes some kind of conflict.  7/29/15
\usepackage{empheq}
\usepackage{enumitem}
\usepackage{hyperref}
\usepackage{hhline}
\usepackage{numprint}
\usepackage{pifont}
\usepackage{xspace}
\usepackage{placeins}
\usepackage{graphicx}
\usepackage{mwe}
\usepackage{setspace}
\usepackage{multirow}
\usepackage{tabularx,booktabs}
\usepackage{hhline}
\usepackage{booktabs}
\usepackage{changepage}
\usepackage{amsmath}
\usepackage{lineno}

\usepackage{slashbox}

\newcommand{\gigalens}{\texttt{GIGA-Lens}\xspace}
\newcommand{\stein}{\textcolor{red}}
\newcommand{\our}{\textcolor{blue}}

\newcommand{\revs}{\textcolor{black}}

\newcommand{\ho}{\ensuremath{H_0}\xspace}

\newcommand{\overlapstein}{398\xspace}

\newcommand{\steinourDs}{148\xspace}

\newcommand{\lenstrain}{1961\xspace}

\newcommand{\lenstot}{1895\xspace}
\newcommand{\lensprevknown}{125\xspace}
\newcommand{\lensrecent}{258\xspace}
\newcommand{\lensA}{115\xspace}
\newcommand{\lensB}{526\xspace}

\newcommand{\lensC}{1254\xspace}

\newcommand{\newlensA}{51\xspace}
\newcommand{\newlensB}{383\xspace}
\newcommand{\newlensC}{1079\xspace}

\newcommand{\lenstotnew}{1512\xspace}
\newcommand{\newAs}{53\xspace}
\newcommand{\newBs}{384\xspace}
\newcommand{\newCs}{1075\xspace}

\newcommand{\grandtotnew}{3057\xspace}

\newcommand{\steinAs}{404\xspace}

\newcommand{\deploysize}{45 million\xspace}

\newcommand{\tractor}{\textit{The Tractor}\xspace}
\newcommand{\twopr}{^{\prime \prime}}

\newcolumntype{Y}{>{\centering\arraybackslash}X}

\submitjournal{ApJS}

\shorttitle{Strong Lenses in DESI Legacy Surveys DR9}
\shortauthors{Storfer, Huang, Gu, Sheu, Banka et al.}

\begin{document}
%\title{Discovering 1000 New Strong Gravitational Lens Candidates in the DESI Legacy Imaging Surveys}
\title{New Strong Gravitational Lenses from the DESI Legacy Imaging Surveys Data Release 9}

%%%% DECaLS Imaging Survey

\correspondingauthor{Christopher Storfer, Xiaosheng Huang}
\email{cstorfer@hawaii.edu, xhuang22@usfca.edu}

% \correspondingauthor{Xiaosheng Huang}
% \email{xhuang22@usfca.edu}
%, august.muench@aas.org}

\author[0000-0002-0385-0014]{C.~Storfer}
\affiliation{Institute for Astronomy, University of Hawaii, Honolulu, HI 96822-1897}
\affiliation{Physics Division, Lawrence Berkeley National Laboratory, 1 Cyclotron Road, Berkeley, CA, 94720}

\author[0000-0001-8156-0330]{X.~Huang}
\affiliation{Department of Physics \& Astronomy, University of San Francisco, San Francisco, CA 94117-1080}

\author[0000-0003-2748-7333]{A.~Gu}
\affiliation{Department of Physics, University of California, Berkeley, Berkeley, CA 94720}
\affiliation{Department of Electrical Engineering \& Computer Sciences, University of California, Berkeley, Berkeley, CA 94720}

\author[0000-0003-1889-0227]{W.~Sheu}
\affiliation{Physics Division, Lawrence Berkeley National Laboratory, 1 Cyclotron Road, Berkeley, CA, 94720}
\affiliation{Department of Physics \& Astronomy, University of California, Los Angeles
430 Portola Plaza, Los Angeles, CA 90095}

% \author{S.~Bailey}
% \affiliation{Physics Division, Lawrence Berkeley National Laboratory, 1 Cyclotron Road, Berkeley, CA, 94720}

% \author[0000-0002-5042-5088]{S.~Bailey}
% \affiliation{Physics Division, Lawrence Berkeley National Laboratory, 1 Cyclotron Road, Berkeley, CA, 94720}

\author{S.~Banka}
\affiliation{Department of Electrical Engineering \& Computer Sciences, University of California, Berkeley, Berkeley, CA 94720}

\author[0000-0002-4928-4003]{A.~Dey}
\affiliation{NSF's National Optical-Infrared Astronomy Research Laboratory, 950 N. Cherry Ave., Tucson, AZ 85719}

\author{A.~Jain}
\affiliation{Department of Electrical Engineering \& Computer Sciences, University of California, Berkeley, Berkeley, CA 94720}

\author[0000-0001-9802-362X]{J.~Kwon}
\affiliation{\revs{Department of Physics, University of California, Santa Barbara, Santa Barbara, CA 93106}}

\author[0000-0002-1172-0754]{D.~Lang}
\affiliation{Perimeter Institute for Theoretical Physics, Waterloo, ON N2L 2Y5, Canada}
\affiliation{\revs{Department of Physics and Astronomy, University of Waterloo, Waterloo, ON N2L 3G1, Canada}}

\author{V.~Lee}
\affiliation{Department of Physics, University of California, Berkeley, Berkeley, CA 94720}

\author[0000-0002-1125-7384]{A.~Meisner}
\affiliation{NSF's National Optical-Infrared Astronomy Research Laboratory, 950 N. Cherry Ave., Tucson, AZ 85719}

\author[0000-0002-2733-4559]{J.~Moustakas}
\affiliation{Department of Physics and Astronomy, Siena College, 515 Loudon Rd., Loudonville, NY 12211}

\author{A.D.~Myers}
\affiliation{Department of Physics \& Astronomy, University of Wyoming, 1000 E. University, Dept 3905, Laramie, WY 82071}

\author{S.~Tabares-Tarquinio}
\affiliation{Physics Division, Lawrence Berkeley National Laboratory, 1 Cyclotron Road, Berkeley, CA, 94720}
\affiliation{Department of Physics \& Astronomy, University of San Francisco, San Francisco, CA 94117-1080}

\author[0000-0002-3569-7421]{E.F.~Schlafly}
\affiliation{Lawrence Livermore National Laboratory,
7000 East Ave., Livermore, CA 94550-9234}

\author[0000-0002-5042-5088]{D.J.~Schlegel}
\affiliation{Physics Division, Lawrence Berkeley National Laboratory, 1 Cyclotron Road, Berkeley, CA, 94720}

%\collaboration{(The DESI Collaboration)}

\begin{abstract}
We have conducted a search for strong gravitational lensing systems in the Dark Energy Spectroscopic Instrument (DESI) Legacy Imaging Surveys  Data Release~9. 
%We have successfully mined for strong lenses in earlier data releases of these surveys \citep[][Paper~I \& II, respectively]{huang2020a, huang2021a}. 
This is the third paper in a series  \citep[following][Paper~I \& II, respectively]{huang2020a, huang2021a}.  
These surveys together cover $\sim$ 19,000~deg$^2$ 
visible from the northern hemisphere, reaching a \revs{$z$-band} AB
magnitude of $\sim 22.5$. 
We use a deep
%%We search in the Dark Energy Spectroscopic Instrument Legacy Imaging Surveys for new strong lensing systems by using deep 
residual neural network, 
trained on a compilation of  
known lensing systems and \revs{high grade} candidates as well as  
non-lenses in the same footprint.
After applying our trained neural network to the survey data, we visually inspect and rank images 
with probabilities 
above a threshold.
We have found \lenstot lens candidates, of which \textit{\lenstotnew are identified for the first time}.
Combining the discoveries from this work with those from Paper~I (335) and II (1210), we have discovered a total of
\grandtotnew \textit{new} candidates in the Legacy Surveys.
%, the largest number of discoveries from any survey to-date.

\end{abstract}

% \section{New Material}
% \label{sec:new}
% \input{new-DR9}
\vspace{0.5in}
\section{Introduction}
\label{sec:intro}
Strong gravitational lensing systems are a powerful tool for cosmology. They have been used to study how dark matter is distributed in galaxies and clusters \citep[e.g.,][]{kochanek1991a, koopmans2002a, bolton2006a,koopmans2006a,bolton2008a, bradac2008a, huang2009a,  grillo2015a, tessore2016a,shu2016a,shu2017a},  
and are uniquely suited to probe substructure in cluster and galaxy scale lenses, 
as well as line-of-sight low-mass halos and test the predictions of the cold dark matter (CDM) model beyond the local universe \citep[e.g.,][]{vegetti2009a, vegetti2010b, vegetti2012a, hezaveh2016a, ritondale2019a, meneghetti2020a, cagansengul2021a,wagner-carena2022}.
%,sengul2021}. 
Multiply lensed supernovae (SNe) are ideal for measuring time delays and \ho because of their well-characterized light curves, and in the case of Type~Ia, 
with the added benefit of standardizable luminosity \citep{refsdal1964a, treu2010a, oguri2010a}, provided microlensing can be accurately characterized \citep{yahalomi2017a}.
%,foster2018}. 
Furthermore, lens models can be constructed after the SNe have faded \citep{ding2021a}.
In the last decade, strongly lensed supernovae,  both core-collapse \citep{kelly2015a, rodney2016a} and Type~Ia \citep{quimby2014a, goobar2017a, rodney2021a,goobar2022a,pierel2022a,chen2022a} have been discovered. Very recently, \citet{sheu2023a} conducted a retrospective search for strongly lensed supernovae in the Dark Energy Spectroscopic Instrument (DESI) Legacy Imaging Surveys (Dey et al. 2019), and found seven promising candidates.
Time-delay \ho measurements from multiply imaged supernovae \citep[e.g.,][]{goldstein2017a, shu2018a, goldstein2018a, goldstein2018b, pierel2019a, suyu2020a, huber2021a}, 
combined with measurements from distance ladders \citep[e.g.,][]{riess2019, freedman2019a, freedman2020a,riess2021a} and lensed quasars \citep[e.g.,][]{suyu2010a, suyu2013a, treu2016a, bonvin2017a, wong2019a,millon2020a,birrer2020a}, can be an important test of the tension  between \ho measured locally and the value inferred from the Cosmic Microwave Background \citep[CMB;][]{planck2020}.
In addition, magnified (but not multiply-lensed) SNe~Ia were identified \citep{patel2014a, nordin2014a, rubin2018a} and used to test the lens models.
Finally strong lensing systems can be used to constrain the properties of dark energy \citep[e.g.,][]{treu2010a, linder2016,sharma2022a}.
%-- say multiple source plane systems will be discovered.  With thousands of lensing systems found, some are bound to be revealed to have another lensed source with high resolution imaging with enough depth 

The introduction of neural networks to identify gravitational lens candidates in imaging surveys has 
%ushered in a new epoch for lens discovery 
been transformative \citep[e.g.,][]{jacobs2017a, metcalf2018a, jacobs2019a, jacobs2019b, canameras2020a,canameras2021a}. 
%In our recent work, we discovered over 1500  
%335 \citep{huang2020a} + 1312 \citep{huang2020b} 
%new strong lenses \citep{huang2020a, huang2021a}
%Paper~I \& II
%in the Dark Energy Spectroscopic Instrument (DESI) Legacy Imaging Surveys by using residual neural networks. 
In our recent work, we discovered over 1500 new strong lenses \citep[][Paper~I and II, respectively]{huang2020a, huang2021a} in the DESI Legacy Imaging Surveys \citep{dey2019a} by using residual neural networks trained on observed images.

In this paper, we present results from our third search for strong lenses in the Legacy Surveys, using Data Release 9 (DR9).
%and improves considerably upon existing methods.  
We provide an overview of the observations in \S\ref{sec:observations}.
In \S\ref{sec:model-train}, we describe the construction of the training sample and our neural network model.  
New lens candidates are presented in \S\ref{sec:results}.
We discuss our discoveries in \S\ref{sec:discussion} and conclude in \S\ref{sec:conclusion}.

% \FloatBarrier
\section{Observations}
\label{sec:observations}
The Legacy Surveys is composed of three surveys: the Dark Energy Camera Legacy Survey (DECaLS), the Beijing-Arizona Sky Survey (BASS), and the Mayall $z$-band Legacy Survey (MzLS). 
DECaLS is observed by the Dark Energy Camera \citep[DECam;][]{flaugher2015a} on the 4-m Blanco telescope, which covers $\sim 9000$~deg$^2$ of the sky in the range of $-18^{\circ} \lesssim \delta \lesssim +32^{\circ}$. 
BASS/MzLS are observed in the $g$ and $r$ bands by the 90Prime camera \citep[][]{williams2004a} on the Bok 2.3-m telescope and in the $z$-band by the Mosaic3 camera \citep[][]{dey2016a} on the 4-meter Mayall telescope. 
Together BASS/MzLS cover the same $\sim5000$~deg$^2$ of the northern subregion of the Legacy Surveys.
%In its entirety, the Legacy Surveys covers $\sim14,000$ deg$^2$ of extragalactic sky in three optical bands (g, $r$, $z$).
Data Release 9 (DR9)
% Dey, Schlegel et al. 2022, in prep) 
contains additional DECam data reprocessed from the Dark Energy Survey \citep[DES;][]{des2016} for $\delta\lesssim-18^{\circ}$.
This provides an additional $\sim 5000~$deg$^2$, resulting in a total footprint of \revs{$\sim19,000$~deg$^2$}. 
The DECam surveys will hereafter be referred to in \revs{their} entirety as DECaLS, 
within which we distinguish DES and non-DES regions. 
% The combined footprint is divided into two contiguous regions separated by the Galactic plane. 
The Legacy Surveys is imaged with a median $5\sigma$ \revs{PSF} depth of 22.5 AB mag in $z$-band \revs{(22.6 AB mag for MzLS).}
The FWHMs for the delivered images are: $1.29\twopr$ ($g$), 1.18$\twopr$ ($r$), and 1.11$\twopr$ ($z$)  for DECaLS; $1.61\twopr$ ($g$) and $1.47\twopr$ ($r$) for BASS; and $1.01\twopr$ ($z$) for MzLS.  

\textit{The Tractor} \citep[][]{lang2016a} is a forward modeling algorithm that performs probabilistic astronomical source detection and typing and constructs the source catalogs for the Legacy Surveys. 
Source extraction is done on pixel-level data, taking as input the individual images from multiple exposures in multiple bands, with different seeing in each. 
\revs{\textit{The Tractor} treats the fitting process as a $\chi^2$ minimization problem. \revs{A detected source is retained if the initial penalized $\chi^2$ is improved by 25\footnote{For more details, see \href{https://www.legacysurvey.org/dr9/description/}{https://www.legacysurvey.org/dr9/description/}}.}} 
\textit{The Tractor} models detected sources as the better of either a point source (``{\tt PSF}'') or round exponential (``{\tt REX}'') galaxy.
\revs{A detected source} can be further classified as de~Vaucouleurs ({\tt DEV}; S\'ersic index $n=4$) or exponential ({\tt EXP}; $n=1$) profile \revs{over {\tt REX}/{\tt PSF} if such a fit improves the} $\chi^2$ by 9. 
The classification becomes a S\'ersic profile ({\tt SER}) with an improvement in the $\chi^2$ by another 9 over {\tt DEV}/{\tt EXP}. 
% Thus, {\tt DEV}/{\tt EXP} and {\tt SER} classifications are $>5.8\sigma$ and $>6.5\sigma$ detections, respectively. 
Earlier data releases included a composite ({\tt COMP}) profile,
which is no longer \revs{fit for} in DR9.

Figure~\ref{fig:dr9-footprint} shows the     
depth map of $z$-band observations in the Legacy Surveys DR9 by plotting the depth of objects typed as {\tt SER} with $z$~$<~20.0$~mag. 
{\tt SER} is the most common galaxy type in this magnitude regime. Table~\ref{tab:objects_z20} shows the total counts for each galaxy type ({\tt SER}, {\tt DEV}, {\tt REX}, and {\tt EXP}) with $z$~$<~20.0$~mag for both the BASS/MzLS and DECaLS regions. 

\begin{deluxetable*}{lcccccccc}[h]
\tablewidth{0pt}
\tabletypesize{\scriptsize}
\tablecaption{Object Counts\label{tab:objects_z20}}
\tablehead{\colhead{\tractor Type} & \colhead{SER} & & \colhead{DEV} & &  \colhead{REX} & \colhead{EXP} & \colhead{Total by Region}}
\startdata
  BASS/MzLS &$3,095,371$ & & $2,942,176$ &  & $2,806,991$ & $1,963,786$ & $10,808,324$\\
  DECaLS  &$15,257,135$ & & $7,223,053$ & & $7,569,493$ & $4,404,247$ & $34,453,928$\\
 \hline
 Total by Type &  $18,352,506$ & & $10,165,229$ & & $10,376,484$ & $6,368,033$ & $\textbf{45,262,252}$ \\ 
\enddata
\end{deluxetable*}

\begin{minipage}{\linewidth}
\makebox[\linewidth]{
  \includegraphics[keepaspectratio=true,scale=0.4]{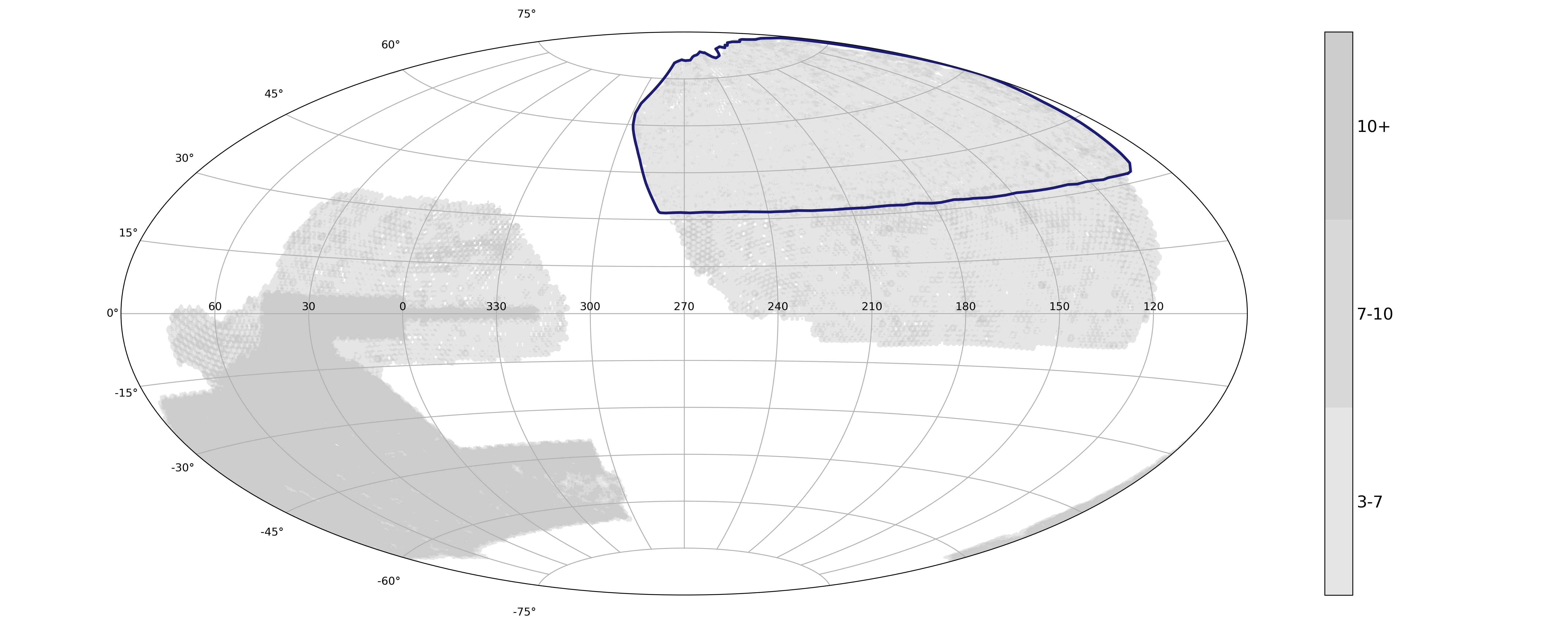}}
  % \captionsetup{font=footnotesize}
\captionof{figure}{\footnotesize The DESI Legacy Imaging Surveys Data Release 9 footprint in an equal-area Aitoff projection in equatorial coordinates. 
The blue border approximately outlines the MzLS/BASS region and the rest is DECaLS. 
The gray shading indicates the $z$ band depth. 
The light shade represents objects observed with between three and seven passes, medium for between seven and ten, and dark for more than ten. 
The contiguous region of $>$10 passes represents the original DES footprint. Note that, prior to DR9, the DES region of DECaLS at $\delta\lesssim -32^{\circ}$ was not entirely reprocessed by the Legacy Surveys imaging team. 
The footprint of the Legacy Surveys DR9 now covers $\sim$
19,000 deg$^2$.
}
\label{fig:dr9-footprint}
\end{minipage}

\section{The Training Sample and Neural Network Model}
\label{sec:model-train}
Convolutional neural networks (CNNs) have been very successful in identifying strong gravitational lenses in simulated and real imaging data.
%With the number of known candidate lenses now in the range of five to ten thousand, construction of a large sample of high quality lenses to train a CNN is feasible (although still much smaller in size than simulated samples). 
As in Papers~I and II, we continue to use observed images of both lenses and non-lenses for training. 
In what follows, we describe our training sample (\S\ref{sec:train}) and neural network model (\S\ref{sec:model}).
%, an improvement on the neural network architecture reported in \citet[][L18]{lanusse2018a}, which was adopted in H20 and H21. This "shielded" model was also used in a supplementary role in H21. 

\subsection{Training Sample}\label{sec:train}

The training sample contains \lenstrain lenses, 
the majority of which come from Paper~I and II 
%(1404)
with the rest selected from \citet[]{carrasco2017a, diehl2017a, pourrahmani2018a, sonnenfeld2018a, wong2018a, jacobs2017a, jacobs2019a, jacobs2019b,petrillo2019a,canameras2020a}.

\revs{To date, $\mathcal{O}(100)$ systems \revs{have been} confirmed via high-resolution imaging and/or spectroscopy \citep[e.g., MasterLens;][]{moustakas2012a}. Thus, a large majority of the lens candidates in the training sample are not confirmed. Some of these candidates were found by searches conducted by other groups in various imaging surveys, using different search methods and criteria. To select the lenses for our training sample, we apply a uniform set of criteria (similar to those in Paper I and II) based solely on the Legacy Surveys images for these systems, regardless of where they were originally discovered. The locations of the selected systems are shown in Figure~\ref{fig:train-lens-map}.} 
% The locations of the lenses on the sky are shown in Figure~\ref{fig:train-lens-map}.

%because we are deploying on LS images

%%%%%new lenses only recently being confirmed spectroscopically tran, glazebrook+huang hst overall known systems confirmed with spec/imag is several hundred therefore vast majority of these sytems NOT confirmed

%%%this represents BULK of known lenses

% 

 % \revs{A large majority of the lenses in the training sample are not confirmed, either by high resolution imaging or spectroscopically. Additionally, some of these lenses were discovered in various imaging surveys (e.g. Hyper Suprime Cam; HSC). Thus the lenses in the training sample were vetted based almost entirely on their quality in the Legacy Surveys images. Given the limitation on the number known lens candidates, ensuring the quality of the lenses in the training sample is crucial.}

\revs{As in our previous searches, we apply cuts} on the type ({\tt SER}, {\tt DEV}, or {\tt REX}), magnitude ($z<20.0$~mag), and depth ($\geq$3 passes in $g$, $r$, and $z$ bands) of the objects included in the training sample. 
\revs{We find that the inclusion of objects fainter than \revs{$z$~$=$~20.0 mag} results in diminishing returns for the search.}
% \revs{The cut on the number of passes was made to ensure image quality, according to the standard set by the Legacy Surveys.}
\revs{These same cuts are applied to images in the training sample and for deployment} \revs{with the exception of images centered on {\tt EXP} objects, which are included in deployment, but not in the training sample. To-date there are relatively few \revs{lens galaxies} typed as {\tt EXP}. However, low surface brightness, extended lensed source images are sometimes typed as {\tt EXP}.}
% Inclusion of images centered on {\tt EXP} objects increases the chance that they are selected by the neural network; this is especially true when no {\tt SER}, {\tt DEV}, or {\tt REX} galaxies which meet our cuts are included in an image.

\begin{minipage}{\linewidth}
\makebox[\linewidth]{
  \includegraphics[keepaspectratio=true,scale=0.4]{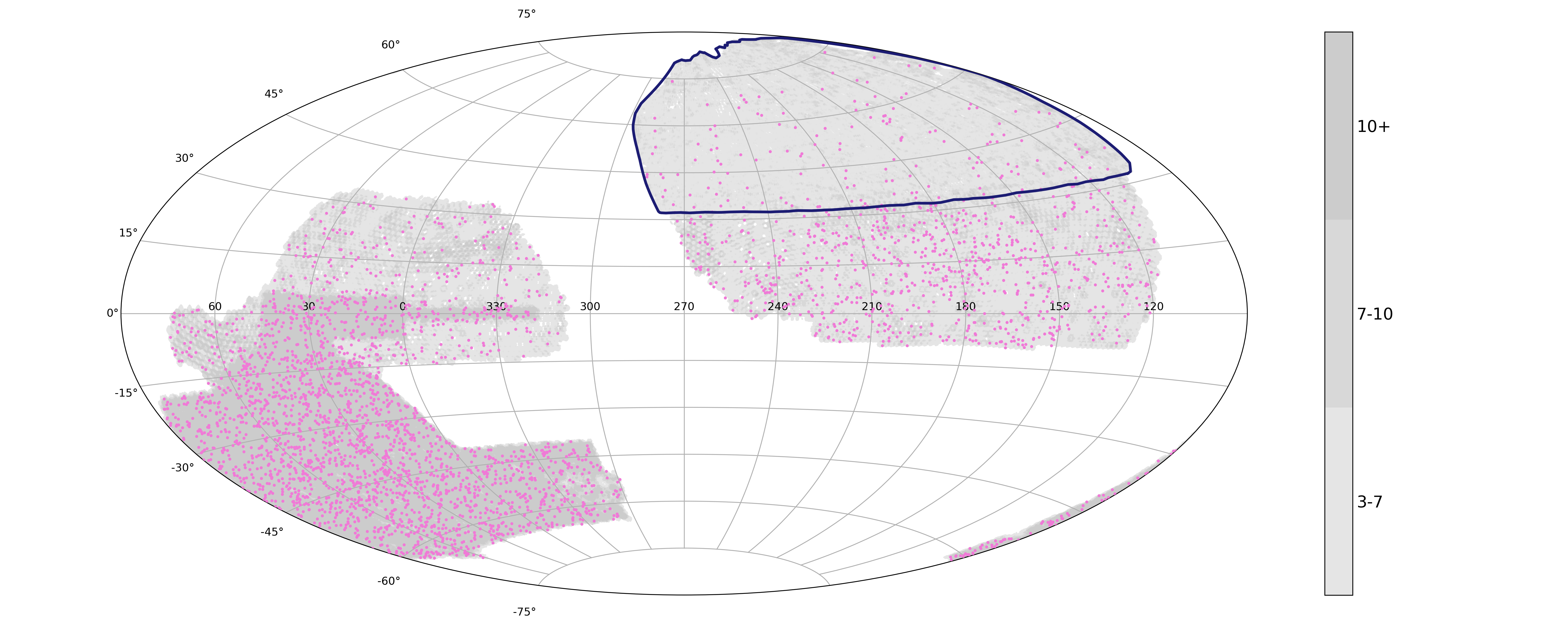}}
    % \captionsetup{font=footnotesize}
\captionof{figure}{\footnotesize The \revs{1961} lenses included in the training sample over the depth map of the Legacy Surveys DR9 shown in Figure~\ref{fig:dr9-footprint}.}
\label{fig:train-lens-map}
\end{minipage}

%Selection of non-lenses in H21 was done by taking a random sample of objects within DECaLS, representative of the objects the neural network was deployed on. 
% The sample of non-lenses which is statistically representative of the lenses in the training sample. 
For the selection of non-lenses, for each type ({\tt SER}, {\tt DEV}, {\tt REX}) and region (DECaLS, BASS/MzLS), 
we bin \revs{by} the number of passes in $z$-band. 
%and determine the percentage
%fractional representation in each bin. 
We then select non-lenses randomly, but keep the proportionality to the lenses the same in each bin ($\sim$~33:1). 
This is to prevent potential bias by the neural net based on the number of passes (see Paper~II). 
%or region.
%preferential selection of images with a high number of passes by the neural net. 
%This issue was noticed in H20 and largely corrected in H21. 
% Figure~\ref{fig:lens-nlens-dist} shows the fraction of lenses and non-lenses in the training sample binned by the number of passes in the z-band. The distributions match nearly exactly with a bin width of two. 
% We maintain the same lens to non-lens ratio as in Paper~II ($\sim$1/33), 
This results in a total of 64,584 non-lenses in the training sample. 
%These non-lenses were inspected by co-authors C.S., X.H., A.G., and W.S. to remove potential contaminants (i.e., lenses). 

%\begin{minipage}{\linewidth}
%\makebox[\linewidth]{
%\includegraphics[keepaspectratio=true,scale=0.4]{train_lens_nlens_nobs_v2.png}}
%\captionof{figure}{The distribution of lenses (purple) and non-lenses (orange), binned by the number of passes in the $z$-band.}
%\label{fig:lens-nlens-dist}
%\end{minipage}

\subsection{The Neural Network Model}\label{sec:model}

The ``shielded'' residual network (ResNet) model, as described in Paper~II, is an improvement upon the neural network architecture presented in \citet[][L18]{lanusse2018a}.
The ``shielding'' layers \citep{szegedy2014a} perform 1$\times$1 convolutions that reduce the dimensionality of the output from each ResNet block.
This modification improved performance and shortened training time. 
During the training, the ResNet attempts to minimize the cross entropy loss function:
\begin{linenomath*}
\begin{equation}
\label{eqn:loss}
    \displaystyle\revs{\mathcal{L}_{CE} =} -\sum_{i=1}^{N} y_i \log \hat{y}_i+(1-y_i) \log (1-\hat{y}_i)
\end{equation}
\end{linenomath*}

\noindent
where $y_i$ is label for the $i$th image (1 for lens and 0 for non-lens), $\hat{y}_i \in [0,1]$ is the model predicted probability\revs{, and $N$ is the number of images in one training step (the same as batch size, given below)}.
In the training process, the ``shielded'' model outperformed the L18 model and thus we decide to exclusively use it in this work. 
We use a 7:3 split of the training sample to create the training and validation sets, respectively.  
We use an image cutout size of 101~pixel$\times$101~pixel \revs{(which translates to $\sim$ 26$^{\prime\prime}\times$26$^{\prime\prime}$)}, 
a batch size of 128, an initial learning rate of $5\times10^{–4}$, 
a decay rate of 1/5, and a decay epoch at 80. 
%We find these new hyperparamters to be particularly suitable for the "shielded" model. 
The model is trained on Google Colab\footnote{\href{https://colab.research.google.com/}{https://colab.research.google.com/}} using a graphics processing unit (GPU\revs{; NVIDIA Tesla P100)}.
%(GPU; NVIDIA Tesla P100). 
%The Tesla P100 GPU was accessed using Colab Pro account which also grants 32GB of RAM, allowing for fast training time of our deep neural network on a large training sample. 
We trained our ResNet model for a total of 145 epochs (approximately 5 hours).
\revs{Training beyond this point results in marginal gain in model performance (see Figure~\ref{fig:loss_roc})}\revs{.}
%We use the best-performing model for deployment.
%initially train the model for 120 epochs, saving the best model by setting TensorFlow's \verb|save_best_only=True| which saves only if the validation accuracy is higher than the previously saved model. Taking the best model, saved at epoch 85, we continue training for an additional 60 epochs in an attempt to achieve improved convergence of model metrics. The best model was saved on epoch 133. 
In addition to the cross entropy loss, we further assess the performance of of our trained model by using the receiver operating characteristic (ROC).
The ROC curve shows the True Positive Rate (TPR) vs. the False Positive Rate (FPR), where P(ositive) indicates a lens and N(egative), a non-lens.
With the definitions True Positive (TP) = correctly identified as a lens, 
False Positive (FP) = incorrectly identified as a lens, 
True Negative (TN) = correctly rejected, and False Negative (FN) = incorrectly rejected,
\begin{equation*}
    \rm{TPR} = \frac{\rm{TP}}{\rm{P}} = \frac{\rm{TP}}{\rm{TP} + \rm{FN}}
\end{equation*}

and
\begin{equation*}
    \rm{FPR} = \frac{\rm{FP}}{\rm{N}} = \frac{\rm{FP}}{\rm{FP} + \rm{TN}}
\end{equation*}

Random classifications will result in a diagonal line in this space with an area under the ROC curve (or AUC) equal \revs{to} 0.5,
while for a perfect classifier, AUC = 1.  
The ROC curve for our best-performing model is also shown in Figure~\ref{fig:loss_roc} for the validation set, with an AUC of 0.9997, \revs{an improvement over the training results from Paper II}. 
%This is the lowest loss and the highest AUC of any model this group has trained (a significant improvement over the AUC of 0.992 in H21). 
%This is most likely attributed to the increase in the overall size of the training sample, the improved quality of the lens sample, and the statistically representative nature of the non-lens sample. The non-lens sample being representative of the lens sample (in terms of type, region, and number of passes in z-band) intuitively lead to improved performance during training.

\begin{minipage}{\linewidth}
\makebox[\linewidth]{
  \includegraphics[keepaspectratio=true,scale=0.55]{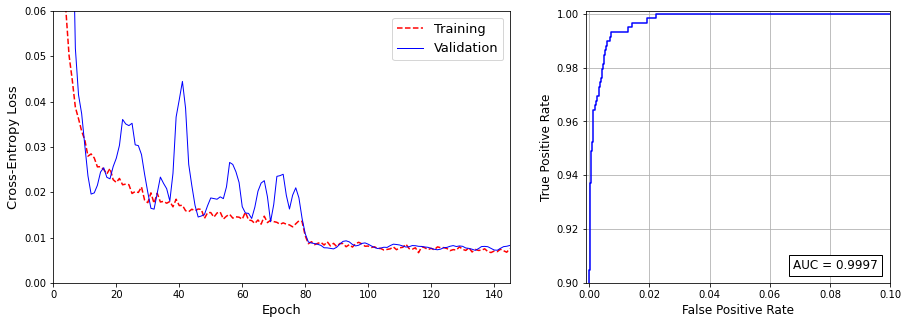}}
  % \captionsetup{font=footnotesize}
\captionof{figure}{\footnotesize Left: the cross-entropy loss for the training and validation sets for 145 epochs. Right: the receiver operating characteristic curve (ROC) for the validation set using the best-performing model. }
\label{fig:loss_roc}
\end{minipage}

\section{Results}
\label{sec:results}
In this section we present the lens candidates we have found in the Legacy Surveys. 
 In \S\ref{sec:dr9-cands}, we present lens candidates found in DR9 (this work).
 In \S\ref{sec:all-cands} we \revs{provide a summary of} all lens candidates discovered by \revs{our} group in the Legacy Surveys DR7, 8, and 9. 
%These candidates include all DR9 galaxy types (SER, DEV, REX, or EXP) in DECaLS and BASS/MzLS. 
 \subsection{Lens Candidates in DR9}\label{sec:dr9-cands}
 In this section we present the strong lens candidates discovered exclusively in Legacy Surveys DR9 from this work. 
 To determine the probability threshold for human inspection,
we consult the precision-recall curve (PRC), 
where  precision = TP/(TP+FP) and recall = TP/(TP + FN), which is the same as TPR (\S~\ref{sec:model-train}).\footnote{\revs{As in Paper II, we recognize the redundancy in terminology.
This results from fairly standard usage 
(e.g., recall or TPR depending on the context) 
and in part from the difference in terminology used in the fields of machine learning and astrophysics (``recall'' or ``completeness'', ``precision'' or ``purity'').}}

Figure~\ref{fig:prc} shows the  precision-recall curve (PRC) for our trained ResNet model for the validation set, with marked probability thresholds. 
% The PRC shows the purity and completeness of the fully trained model on the validation set. The PRC evaluate the fraction of true positives among positive predictions and thus can provide an accurate prediction of future classification performance, especially in the case of a highly asymmetrical, binary classification problem \citep{takaya2015}. 
The distribution of the probability thresholds on our PRC are tightly clustered near the ``shoulder" of the curve, indicating a high level of performance. 
A perfect PRC curve would pass through (1,1) where these thresholds would converge. 
We consult the PRC to determine the probability threshold for inspection
based on a reasonable trade-off between precision (purity) and recall (completeness). 
We select a probability threshold of 0.4, with the number of cutout images above this threshold $\sim$~47,000.  
%51,804, after removing duplicates, 46779
This is a reasonable number to inspect, comparable to those in Paper~I and II. 
These cutouts with probability $\geq$ 0.4 will be referred to as the neural network ``recommendations".

For deployment, 
%in our previous searches, \citet[][Paper~I]{huang2020a} was restricted to elliptical galaxies (those typed as DEV or COMP) in the DECaLS region of the Legacy Surveys. 
%\citet[][Paper~II]{huang2021a} expanded the search to the entirety of the Legacy Surveys footprint and included galaxies typed as REX -- typically rounder, fainter galaxies 
%despite being trained only on DEV and COMP (elliptical) galaxies in DECaLS. 
%In DR9, DEV (elliptical galaxies) and SER profiles are the most common type of objects. 
%This, along with our success in finding lenses with REX profiles, 
even though our ResNet model is trained on {\tt DEV}, {\tt SER} and {\tt REX}, we expand our search to include galaxies with {\tt EXP} profiles (as mentioned in \S\ref{sec:train}). 
% Spiral galaxies are not typically efficient lenses, but we have noticed in the past that lensed arcs can sometimes be classified as {\tt EXP}, as lensed arcs can appear similar to spiral arms. 
% Thus including this type in deployment could potentially yield additional discoveries. 
%Examples of this are show in Figure~\ref{fig:exp-arcs}, the white circles indicate where \tractor has identified an EXP profile. 
%Therefore, in this paper we will search all galaxy types (SER, DEV, REX, and EXP) in the Legacy Surveys DR9.
%\begin{minipage}{\linewidth}{\vspace{.5cm}}
%\makebox[\linewidth]{
%  \includegraphics[keepaspectratio=true,scale=0.45]{exp_arcs.pdf}}
%\captionsetup{font=footnotesize}
%\captionof{figure}{
%Examples of lenses found in Paper~I and II where \textit{The Tractor} has identified a background lensed source as an EXP object (white circle). 
%Lensed arcs, which can sometimes appear similar to spiral arms (first, third and fourth panels), are sometimes typed as EXP.
%This kind of mistyping, however, is not restricted to blue arcs (the second and fifth panels).}
%\label{fig:exp-arcs}
%\end{minipage}
Thus we deploy the trained model on 45~million
%45,262,252 
cutouts with $\geq 3$ passes in $g$, $r$, and $z$-bands and centered on all non-{\tt PSF} objects with $z$-band mag $\leq 20.0$. 
%By far the largest single deployment ever, by our group or other works. 
Subsequent inspection of the resulting recommendations was conducted considering the following criteria (same as in Paper~I and II):
small blue galaxy/galaxies (red galaxies are rare but certainly acceptable) next to a red galaxy/galaxies at the center \revs{that satisfy the following criteria:} 
    \begin{itemize}%[label=$\circ$]
        \item are typically 1 - 5$\twopr$ away   
        \item have low surface brightness
        \item curve toward the red galaxy/galaxies
        \item have counter/multiple images with similar colors (especially in Einstein-cross like configuration)
        \item are elongated (including semi- or nearly full \revs{rings})

    \end{itemize}
\noindent
Typically, most candidates do not have all these characteristics. 
In general, the greater the \revs{number of these characteristics} an image has, the higher they are ranked by humans.

After an initial round of inspections, 
co-authors C.S. and X.H. independently examine all preliminary candidates.
% and independently assign integer scores from 1 to 4. 
% The final candidates are those that have an average score $\geq$2.0 and do not receive a grade $<$2 by either of the inspectors. 
% The average score is then converted to a grade (A ,B, or C). 
% As with Paper~II, 
The grades are assigned as follows: 
\begin{itemize}
    \item Grade~A:  We have a high level of confidence of these candidates.  
    Many of them have one or more prominent arcs, usually blue.
    The rest have one or more clear arclets, sometimes arranged in multiple-image configurations with similar colors (again, typically blue).  However, there are clear cases with red arcs.
    
    \item Grade~B:   
    They have similar characteristics as the Grade A's.  
    \revs{Grade B giant arcs tend to be fainter than those for Grade A}.  
    Likewise, the putative arclets tend to be smaller and/or fainter, or isolated (without counter images).
    
    \item Grade~C:  
    They generally have features that are even fainter and/or smaller than what is typical for Grade~B candidates, but that are nevertheless suggestive of lensed arclets.  
    Counter images are often not present or \revs{hardly} discernible. 
    In a number of cases,
    %, if these are indeed lensing systems, 
    the angular scales of the candidate systems
    %source image(s)} 
    are comparable to or only slightly larger than the seeing.
    Therefore, for some of these candidates, to attain a higher level of certainty, higher spatial resolution, deeper data, or spectroscopic observations would be required.
\end{itemize}

 \begin{minipage}{\linewidth}
\makebox[\linewidth]{
  \includegraphics[keepaspectratio=true,scale=0.5]{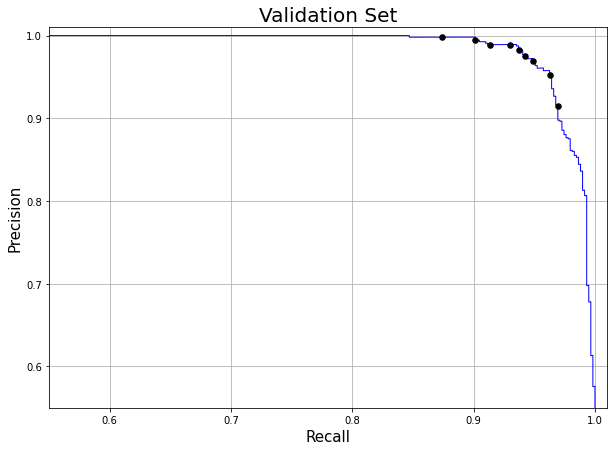}}
  % \captionsetup{font=footnotesize}
\captionof{figure}{\footnotesize The precision-recall curve (PRC) for the validation set from the fully trained neural network. 
The black dots from \revs{right to left} on the curve indicate probability thresholds from 0.1 to 0.9 with a step of 0.1.}
\label{fig:prc}
\end{minipage}

% In total, we deploy our deploy model on $\sim$ 45 million cutouts centered on galaxies of all types with $z$-band mag $\leq$ 20.0 in both DECaLS and BASS/MzLS. 
% %We inspect 51,804 recommendations which received a probability above the threshold of 0.4. 
% We present candidates of all galaxy types (SER, DEV, COMP, REX, EXP) in DECaLS and BASS/MzLS regions. 
 %All candidates were discovered using either the L18 ResNet or the ``shielded'' model. 

We have found \lenstot candidates, with \lensA Grade A's, \lensB B's, and \lensC C's (Table~\ref{tab:shielded-cands}).
% The breakdown by grades is shown in Table~\ref{tab:shielded-cands}.
For the {\tt DEV} and {\tt SER} type galaxies in DECaLS we achieve a purity of $\sim$1 candidate in 23 ResNet recommendations. 
Though it is difficult to do a direct
comparison given the change in \tractor categorization, 
this seems to indicate an improvement over the purity of $\sim$1 in 31 reported in Paper~II for {\tt DEV} and {\tt COMP} in DECaLS.
Certainly, the overall purity of $\sim$ 1 in 25 recommendations is a clear improvement over Paper~II (1 in $\sim 40$).
%By L18 alone, it would be 40000/1000 = 40; and we said "shielded" had comparable human inspection purity.
The purity by type and region is shown in Table~\ref{tab:nn-purity}.
% , along with a summary of all lens candidates in Table~\ref{tab:shielded-cands}. 

\begin{deluxetable*}{lccccccc}[h]
\tablewidth{0pt}
\tabletypesize{\scriptsize}
\tablecaption{Strong Lens Candidates\label{tab:shielded-cands}}
\tablehead{\colhead{Grade} & \colhead{A} & & \colhead{B} & &  \colhead{C} & \colhead{Total by Type}}
\startdata
  SER &68 (64, 4)& &309 (296, 13)&  &650 (630, 20)&1027 (990, 37)\\
  DEV &37 (32, 5)& &113 (108, 5)& &302 (283, 19)&452 (423, 29)\\
  REX &6 (6, 0)& &65 (64, 1)& &224 (219, 5)& 295 (289, 6)\\
  EXP &4 (4, 0)& &39 (34, 5)& &78 (74, 4)& 121 (112, 9)\\
 \hline
 Total by Grade& 115 (106, 9) & &526 (504, 22)& &1254 (1206, 48) & 1895 (1814, 81)\\ 
\enddata
\tablecomments{The numbers shown in the parenthesis correspond to the totals in DECaLS and BASS/MzLS (separated by a comma), respectively.}
\end{deluxetable*}

\vspace{-0.5in}
%\begin{minipage}{\linewidth}
%\makebox[\linewidth]{
\begin{deluxetable*}{lcccccccc}[h]
\tablewidth{0pt}
\tabletypesize{\scriptsize}
\tablecaption{Purity for Our ResNet Model\label{tab:nn-purity}}
\tablehead{\colhead{\tractor Type} & \colhead{SER} & & \colhead{DEV} & &  \colhead{REX} & \colhead{EXP} & \colhead{Purity by Region}}
\startdata
DECaLS  &23& &21&&25& 38& 24\\
BASS/MzLS &29& &34&  &128&51&40\\
 \hline
 Purity by Type&  23& &22& &27 &39&25 \\ 
\enddata
\tablecomments{Purities are shown as the number of ResNet recommendations inspected in order to find a lens candidate.}
\end{deluxetable*}
%}
%\end{minipage}

\newpage

Of these, \lensprevknown were previously known lenses or candidates, and \lensrecent were found in recent publications, the majority of which came from \citet[][see \S\ref{sec:discussion}]{stein2021a}. 
We therefore report \lenstotnew \textit{new} lens candidates,
with \newAs Grade A's, \newBs B's, and \newCs C's. 
We highlight in Figure~\ref{fig:example-cands} two examples each for five categories of new strong lens candidates that we have discovered.
The positions of all new candidates on the sky are shown in Figure~\ref{fig:cand_map_dr9}.

% \newpage

% Among the 25 candidates shown, 
% four have a average human inspection score of 2.5,
% and therefore are given a C grade, 
% but they are nevertheless very likely lensing candidates.
\vspace{0.1in}

 \begin{minipage}{\linewidth}
\makebox[\linewidth]{
  \includegraphics[keepaspectratio=true,scale=0.35]{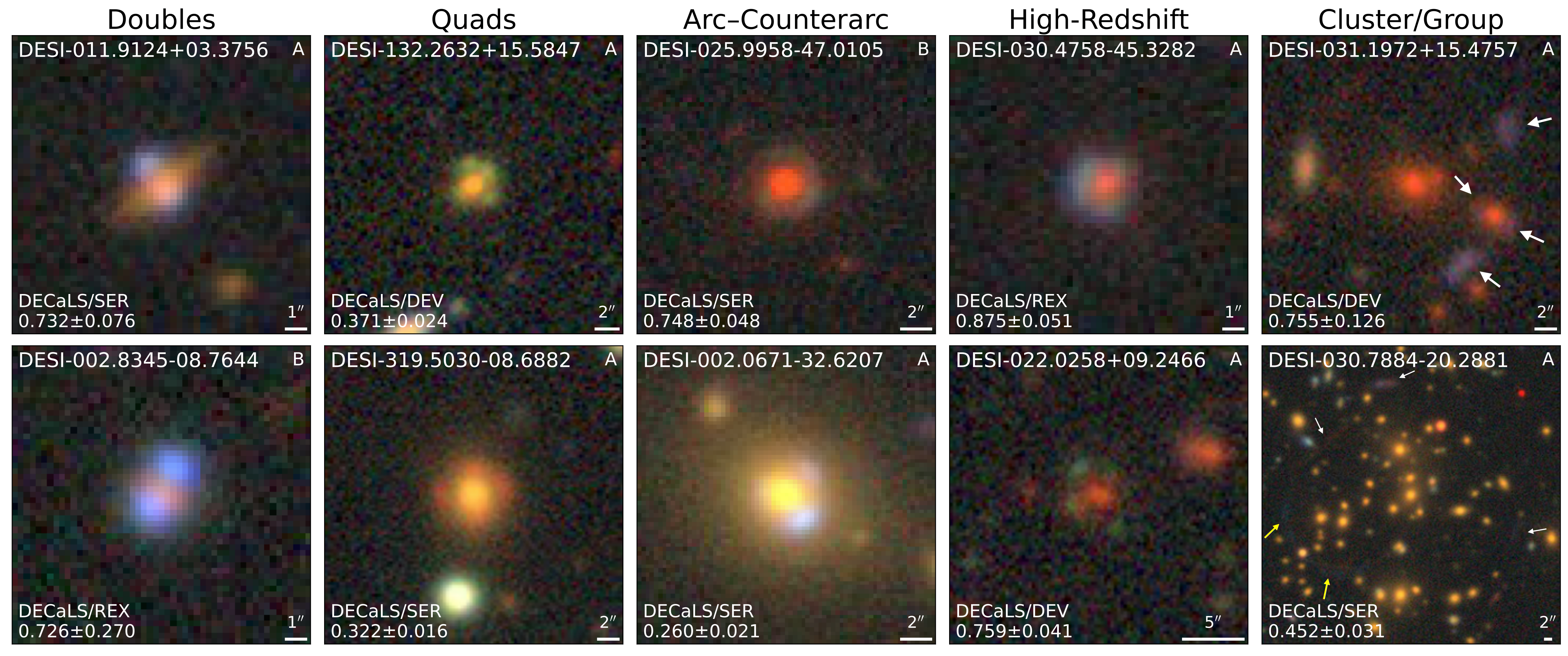}}
  % \captionsetup{font=footnotesize}
\captionof{figure}{\footnotesize Ten of the \lenstotnew new strong lens candidates discovered in this work. 
The naming convention is RA and Dec in decimal format.
\revs{For all images, north is up, and east to the left.}
Top right corner of each image indicates the grade;
bottom left corner, the region/\tractor type, and the photometric redshift of the putative lens ($z_d$).
% and bottom right, the redshift (spectroscopic: SDSS (four decimal places), photometric (three decimal places with uncertainty): \citet{zhou2020a}).
\textbf{First Column:} doubly lensed systems.
\textbf{Second Column:} quadruply lensed systems.
% large arcs. 
\textbf{Third column:} arc-counterarc systems, each of which has a large arc with a smaller counterarc on the opposite side of the lens. 
\textbf{Fourth Column:} high-redshift lensing systems, 
each of which has quadruple images, with two of them merging into an arc.
These eight systems all have a single galaxy as the lens.
%And yet, while the two doubly lensed systems have around $1 \twopr$ Einstein radii, what is considered to be typical for galaxy scale lenses,
%the rest of the systems all have $\gtrsim 2\twopr$ Einstein radii.
\textbf{Fifth Column:} group/cluster lensing systems. \revs{Putative arcs are indicated with white arrows.}
For DESI-031.1951+15.4749, the large lensed arc due to the galaxy group potential is also lensed by a group member, forming two small arcs to its left and right \citep[for another example, see][]{huang2009a}.  
This would enable lensing mass measurement of substructure.
Note that this system also has a high $z_d$ of 0.755.
DESI-030.7884-20.2881 is a cluster lensing system with a faint but spectacularly large blue arc to the lower left \revs{(marked by two yellow arrows to indicate its extent)}.  
There are hints of even fainter arcs around the cluster center. 
High resolution imaging will almost certainly reveal more arcs.
}\label{fig:example-cands}
\end{minipage}

%The first 80 Grade~A candidates (ordered by RA) are shown in Figure~\ref{fig:grade-a-1} and Table~\ref{tab:grade-A}.

\begin{minipage}{\linewidth}
\makebox[\linewidth]{
 \includegraphics[keepaspectratio=true,scale=0.4]{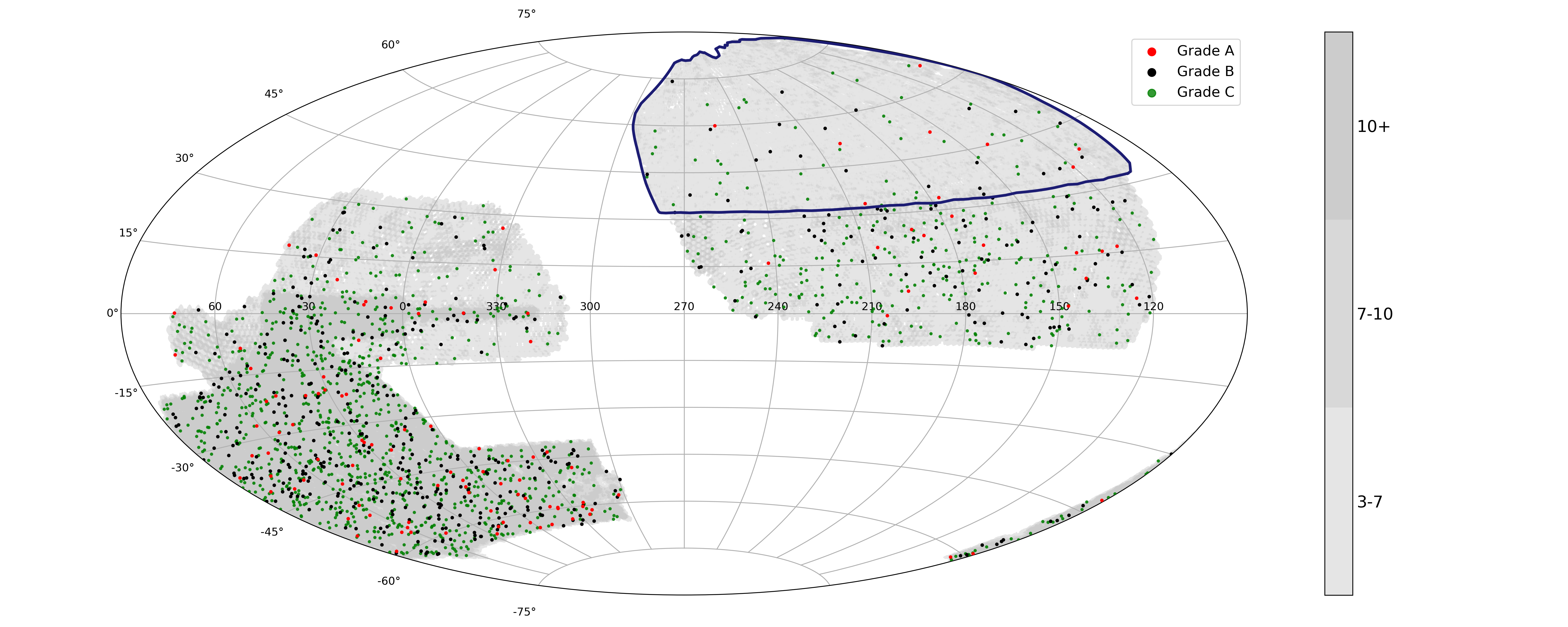}}
% \captionsetup{font=footnotesize}    
\captionof{figure}{\footnotesize The \lenstotnew new candidate lensing systems discovered in this work by grades 
over the depth map of the Legacy Surveys DR9 shown in Figure~\ref{fig:dr9-footprint}.}
\label{fig:cand_map_dr9}
\end{minipage}

% \begin{minipage}{\linewidth}% to keep image and caption on one page
% \makebox[\linewidth]{%        to center the image
% \includegraphics[keepaspectratio=true,scale=0.2]{{grade-a}.pdf}}
% \captionof{figure}{ 
% %\vspace{-70mm}
% \rf{Eighty of t}he \lensA Grade A candidates arranged in ascending RA.  
% Top right corner indicates the average human inspection score with $\Delta$ being the absolute difference;
% bottom left corner, the region and \tractor type (SER, DEV, REX, or EXP);
% and bottom right, the ''shielded" model probability.
% For each image, N is up, and E to the left.
% %For this and all figures that follow:
% Images without rims have a width of 101 pixels ($26.5\twopr$);
% with green rims, 151 pixel ($39.6\twopr$);
% blue rims, 201 pixel ($52.7\twopr$);
% purple rims, 251 pixel ($65.8\twopr$);
% pink rims, 351 pixel ($92.0\twopr$).
% Images with red rims are known lenses \rf{or candidates} but not included in our training sample, 
% with citations given in Table~\ref{tab:grade-A}.
% All \lenstotnew candidates are shown on the project website: \url{https://sites.google.com/usfca.edu/neuralens}.
% %Tables~\ref{tab:grade-a} - \ref{tab:grade-c}}.
% %unless otherwise noted, these are from \citet{jacobs2019b}.
% }\label{fig:grade-a}  
%  \end{minipage}
 
% \input{grade-a.tex}

\subsection{Lens Candidates in DR7, 8, and 9}\label{sec:all-cands}
% We also present the full list of strong lens candidates in the Legacy Surveys, discovered by this group. 
From \revs{our} three searches in the Legacy Surveys, 
we have found a grand total of \grandtotnew \revs{\textit{new}} candidates (Paper~I: 335 in DR7 among {\tt DEV} and {\tt COMP} in DECaLS only, Paper~II: 1210 in DR8, this work: \lenstotnew in DR9). 
The entire catalog of these lenses can be found on our project website\footnote{\href{https://sites.google.com/usfca.edu/neuralens/}{https://sites.google.com/usfca.edu/neuralens/}}. The positions of all candidates on the sky are shown in Figure~\ref{fig:cand_map_all}.
\vspace{-0.5cm}
\begin{minipage}{\linewidth}
\makebox[\linewidth]{
  \includegraphics[keepaspectratio=true,scale=0.4]{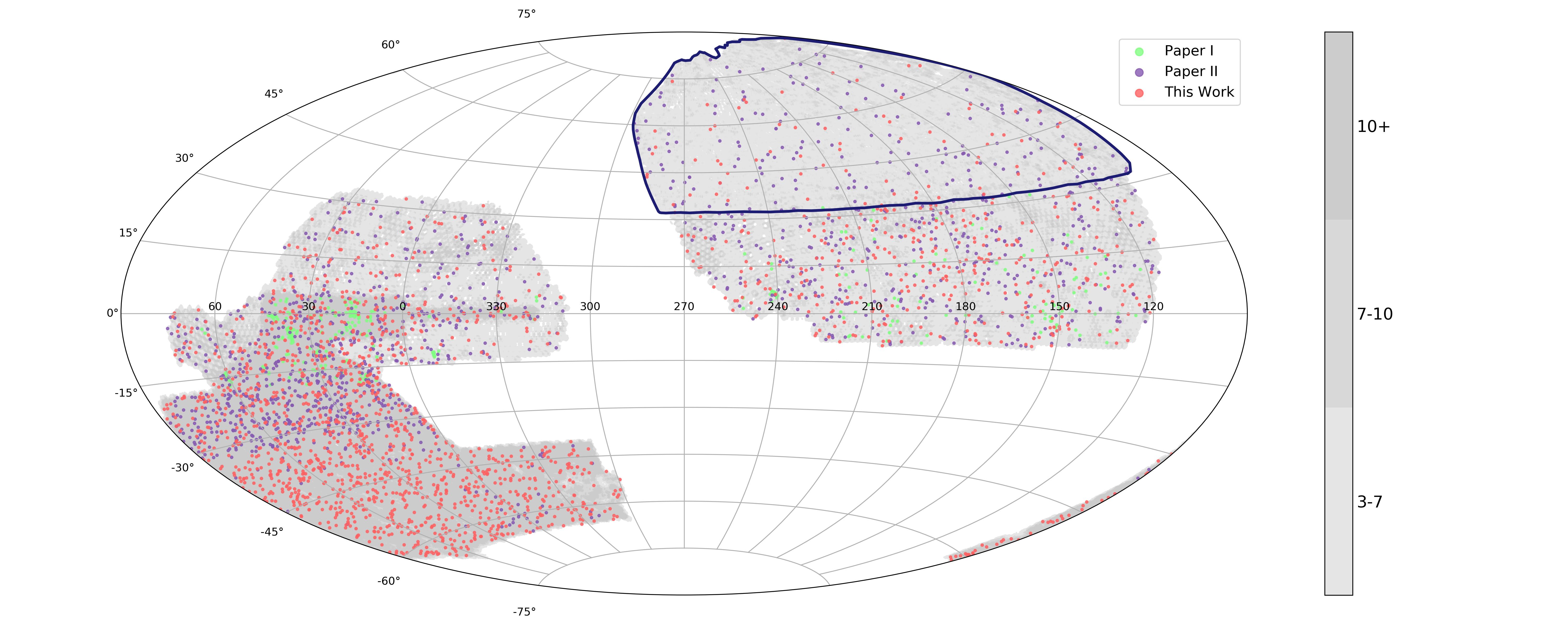}}
% \captionsetup{font=footnotesize}    
\captionof{figure}{\footnotesize All \grandtotnew candidate lensing systems discovered in the Legacy Surveys reported in Paper~I (green), Paper~II (purple), and this work (pink), over the depth map of the Legacy Surveys DR9 shown in Figure~\ref{fig:dr9-footprint}.}
\label{fig:cand_map_all}
\end{minipage}

\newpage
\section{Discussion}\label{sec:discussion}
%This work is the third strong lens search in a series (following Paper~I and II). 
We have trained and deployed on the highly heterogeneous imaging data within the Legacy Surveys, with non-uniform depth and varying image quality across BASS/MzLS and DECaLS, which includes the DES and non-DES regions. 
We show that despite this challenge, 
our neural network can be trained to perform well with a carefully constructed training sample. 
This training sample is larger by a factor of 3 than that in Paper~II, with 
known lenses and lens candidates from both BASS/MzLS and DECaLS, and non-lenses that are selected proportionally in each depth bin for each region. 
%While it is hard to do a direct comparison, Given the difference in object type (SER introduced, and we didn't go EXP last time) and neural network architecture change (Paper II: mixture of L18 and shielded, and here, shielded only),
%in this work the closest comparison comes from measuring efficiency of SER and DEV in DECaLS which was 1 in 24 and 
We achieve an overall ResNet purity of 1 lens candidate in 25 neural net recommendations, a clear improvement over Paper~II (1 in $\sim$40). 
This is competitive with, if not better than, similar searches carried out for other surveys, especially given that the data set we deploy on is the most heterogeneous. 

The DECaLS region of the Legacy Surveys has been mined for lenses several times before
\citep[][Papers~I and II]{diehl2017a, jacobs2019a, jacobs2019b, rojas2021a,  odonnell2022a}.
%(BASS/MzLS by Paper~II; DECaLS non-DES region: Paper~I and II; DECaLS DES region: \citet{jacobs2017a, jacobs2019a, rojas2021a}, Paper~II). 
It is also worth keeping in mind that the Legacy Surveys overlap significantly with SDSS and Pan-STARRS, which contains the entirety of the MzLS/BASS region.
% and the Hyper Suprime-Cam survey, the Kilo-Degree Survey, and CFHTLS are entirely contained within the footprint of the Legacy Surveys, 
Both of these surveys have been mined for strong lenses \citep[e.g.,][]{sonnenfeld2013a, canameras2020a,talbot2021a}. 
Lens searches also have been carried out in the Hyper-Suprime Cam Subaru Strategic Program (HSC SSP), 
Canada–France–Hawaii Telescope Legacy Survey (CFHTLS), and the Kilo-Degree Survey (KiDS), all of which are contained within the Legacy Surveys \citep[e.g.,][]{jacobs2017a, canameras2021a, li2021a}.
Finally, the MzLS/BASS region also significantly overlaps with the Ultraviolet Near Infrared Optical Northern Survey (UNIONS), which \revs{has} been mined for lenses as well \citep{savary2021a}.
%shu2022a?
% \revs{While all surveys differ in depth, seeing, and filter selection, there are a number of lenses which have been co-discovered in two or more of these surveys.}
% With many of the spectacular lensing systems having already been discovered it is not a surprise that the Grade~C candidates make up a slightly larger fraction (0.73) of the reported candidates compared with Paper~II (0.68).
%Nevertheless we point out that 853 of our candidates
%have a human inspection score $\geq 2.5$, all of which are at least likely lensing systems. 
%We deliberately show the numbers for systems with an average score of 2.5 and 2.0.  The former are at least likely candidates.
% (0.73; Paper~I: 0.51, Paper~II: 0.68).  
% But also note the letter grade standards are different.  Human inspections are done by different people
% But significantly, 
And yet, we have succeeded in finding a large number of high quality \textit{new} lenses. 
% \revs{The discoverability of new lenses, by humans alone or with machine learning assistance, of course depends on observation conditions (e.g., depth, seeing, and filter selection) and the criteria for the deployment sample. 
% For machine learning based searches, other factors include the machine learning technique, training sample (e.g., whether they consist of simulations, observed images, or an amalgamation), and  the final human inspection process itself \citep{rojas2023}. These can lead to overlapping and complementary discoveries.}
% results from intra-survey searches 

The photometric redshift distribution of our candidates is shown in Figure~\ref{fig:z-dist}. 
It is largely similar to the redshift distribution of the lenses in the training sample. The average redshift of the new candidates\revs{, however,} is higher by 0.1, at $\sim 0.5$,
%compared with $\sim 0.4$, 
possibly indicating our model's ability to find lenses with a higher redshift distribution than the training sample.

\begin{minipage}{\linewidth}
\makebox[\linewidth]{
  \includegraphics[keepaspectratio=true,scale=0.4]{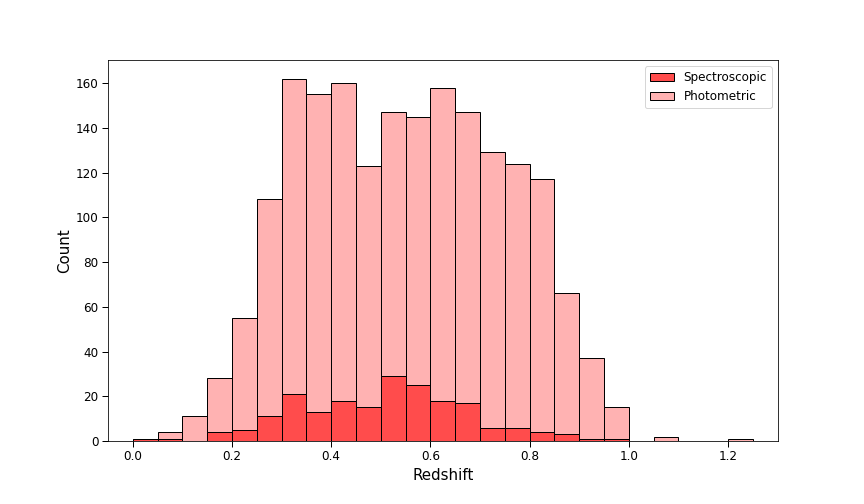}}
  % \captionsetup{font=footnotesize}
\captionof{figure}{\footnotesize Redshift distributions of the candidates found in Legacy Surveys DR9 from SDSS DR17 \revs{(spectroscopic; dark red)} and \citet[][\revs{photometric; light red}]{zhou2020a}}
\label{fig:z-dist}
\end{minipage}

\citet[][S21]{stein2021a} also searched for strong lenses in the DESI Legacy Surveys DR9.
Their approach took two steps.
First they performed a form of self-supervised learning, by applying a CNN encoder on real observed images to minimize contrastive loss in a representation space.
Then, after projecting image cutouts onto this representation space, 
they employed three search strategies: 
similarity search, linear binary classification trained on labeled data, and re-training (``fine-tuning") the self-supervised CNN encoder by incorporating the linear classifier.
% in Sec 4, they said 32*8, or 256 GPU hours.
% (and with much worse efficiency).
%This is the first time the use of self-supervised learning has been used to search for strong lenses. 
As with this work, they searched for lens candidates in all non-PSF objects (typed by \tractor) with \revs{$z$~$<$~20.0~mag}.
They further expanded the search to 
\revs{$z$~$\ge$~20.0~mag}, adding $\sim 17\%$ more candidates.
They reported a total of 1192 strong lens candidates. 
%with high levels of efficiency. 

The candidates reported in S21 were only assigned Grades A and B. 
A total of \overlapstein \revs{($25.7\%$)} candidates from S21 have been identified in our search (limited to \revs{$z$~$<$~20.0~mag}),
which represents approximately
% 316+583+47+131 = 1077
% mag-z < 20: 316+583 = 899 —> 83.5\%
% Assuming same percentage for north:
% mag-z < 20: 1192 * 83.5\% = 995.
40\% of their \revs{$z$~$<$~20.0~mag} candidates.
This includes \steinourDs S21 candidates receiving a grade of D from our visual inspection.
%There are a total of \overlapstein overlapping candidates between S21 and this work. 
% and 21\% of our candidates (for all of which, $z$-mag~$< 20$).
Figure~\ref{fig:grade-comp-stein} shows the distribution of the S21 grades relative to the grades in this work for the overlapping candidates. Table~\ref{tab:stein-comp} shows a two-way comparison.

%(Note that we assigned these grades before S21 was posted on the arXiv). 
%Clearly, our Grade~C systems are overwhelmingly reported as Grade A or B in S21. 

  \vspace{-1cm}

 \begin{minipage}{\linewidth}
\makebox[\linewidth]{
  \includegraphics[keepaspectratio=true,scale=0.5]{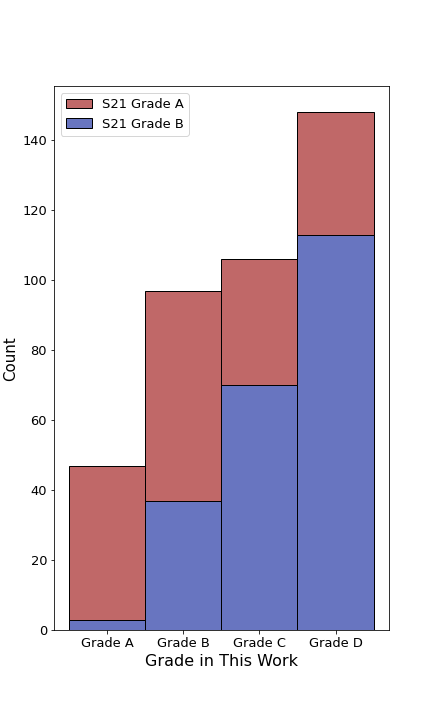}}
  % \captionsetup{font=footnotesize}
\captionof{figure}{\footnotesize The stacked histogram of the distributions of S21 grades shows the comparison with our grades for the overlapping candidates. \revs{Note that grades for the lens candidates reported in this work were assigned prior to S21 being posted on the arXiv.}}
% discovered in S21 and this work. 
% The stacked histogram shows the distribution of S21 grades for the same systems compared to those reported in this work. 
% Grade A and B candidates as reported in S21 are shown in red and blue respectively. 
\label{fig:grade-comp-stein}
\end{minipage}

For the overlapping candidates:
from the red text in Table~\ref{tab:stein-comp}, 
25.8\% of the S21 Grade~A systems are given an A or B grade in this work, 
and a further 8.9\% and 8.7\% of their systems receive a C or D \revs{grade} from our inspection, respectively.  
For the S21 Grade~B systems, 9.6\% are given an A or B grade in our work, 
8.9\% a C grade, and 14.3\% a D grade.
In total, 43.3\% and 28.3\% of the S21 Grade~A and B systems, respectively, are assigned a Grade~D or better from our visual inspection.
Going in the other direction, from the blue text in Table~\ref{tab:stein-comp},
40.9\%, 18.4\%, 8.5\% and 8.0\% of our Grade~A, B, C, and D candidates are identified in S21 (i.e., assigned an A or B grade), respectively.
The agreement is the strongest for mutual Grade~A's.

\revs{Figure~\ref{fig:stein-comp} shows examples of overlapping candidates reported in S21 and this work. 
Grades assigned by S21 (A and B) are the same in each row, and grades from this work are the same in each column (A, B, C, and D). 
While grading criteria may differ between search teams, high quality candidates (grade A and B) are often shared, as shown in the left half of Figure~\ref{fig:stein-comp}.}

\revs{We also provide comparisons for a few example systems with the largest discrepancy in grade: those assigned grade A or B in S21 that were given a grade of C and D in this work.}

\revs{DESI-070.4050-50.4559 (first row, third column) has a faint possible arclet with no visible counter image. There are two other objects of similar brightness but somewhat different color in the cutout (southeast and northwest of the putative lens). They are too faint for us to have high confidence that they are lensed images and thus we assign a grade of C.}

\revs{DESI-153.6301+24.4854 (second row, third column) also shows what appears to be an arclet. While the surface brightness and color are consistent with the expectation of a lensed arc, there is no apparent counter image and the shape of the arc does not conform to the typical morphology (there appears to be a ``kink" in the putative arc, possibly indicating two galaxies in close proximity on the sky). 
It is also possible that the shape of the arc may be due to source structure. Given the limitations of this ground-based image and the absence of a clearly identifiable counter-image, we assign a grade of C.} 

\revs{DESI-066.4365-57.6926 (first row, fourth column) is a type of image that we have seen in our searches a number of times. If it is lensing, this would be a perfect Einstein ring. Since such perfect alignment is very rare even among strong lensing systems, we are generally sceptical. Lensed arcs typically have low surface brightness, whereas this putative ring appears quite bright. The structure seen in the putative ring is, in our experience, consistent with a ring galaxy \citep[e.g,][]{timmis2017a}. 
Furthermore, if this system is indeed a strong lens, the Einstein radius, $\theta_E\sim$~2.8$^{\prime\prime}$. Using a lens redshift of $z_d\sim$~0.4 \citep[photometric redshift;][]{zhou2020a}, and source redshift of $z_s\sim$~0.8 \citep[assumed to be 2$\cdot z_d$; e.g.,][]{sharma2022a}, we can then estimate the mass enclosed within the Einstein radius as well as the velocity dispersion ($\sigma_v$) of the lens. 
Assuming an singular isothermal sphere (SIS) profile, we find the enclosed mass to be $\sim2\times10^{12}$ M$_{\odot}$ and $\sigma_v\sim$~470~km~s$^{-1}$. 
Compared with known lenses in the SLACS program with similar $r$-band magnitude and lens and source redshifts, these values are extraordinarily large. Finally, the galaxy does not appear to be a part of a group.}
% The presumptive foreground galaxy appears to be part of a galaxy group though, not at the center of it. It nevertheless is possible that the arc is produced by the group potential. 
% For example, SLACS~J0903+4116 \citep[][]{bolton2008a} is reported to have a $\sigma_v\sim$~200km~s$^{-1}$ in SDSS DR17, however this system only produces $\theta_E~=~$1.8$^{\prime\prime}$. 
\revs{We therefore assign a grade of D.}

\revs{For DESI-035.4194-46.8752 (second row, fourth column), we consider the putative arc to be more likely a spiral galaxy (perhaps with a high inclination angle), with what appears to be a reddish core and blue spiral arms. In this and other select ambiguous cases, photometric redshfits \citep[][]{zhou2020a} can be helpful. 
The photometric redshift of the putative arc at 0.268$\pm$0.047 relative to that of the putative lens (the orange galaxy at the center of the image) at 0.248$\pm$0.038 is consistent with this not being a lensing system. While photometic redshifts of elliptical galaxies are often reliable, this is not always the case with blue lensed sources. The photometric redshift of the putative arc in DESI-035.4194-46.8752 is centered on the orange core and is likely more reliable. Hence we assign a grade of D.}

\revs{In all four cases, we deem the possibility of lensing to be present (just not high enough to be assigned above C Grade), high resolution image and/or spectroscopic observations are needed to be conclusive.}
% or at least say we always check photo-z in such ambiguous cases.
% Given the two putative lensing galaxies are lined up nearly
% S21 mentioned, they would like others to take a look.
\begin{minipage}{\linewidth}
\vspace{3mm}
\makebox[\linewidth]{
  \includegraphics[keepaspectratio=true,scale=0.3]{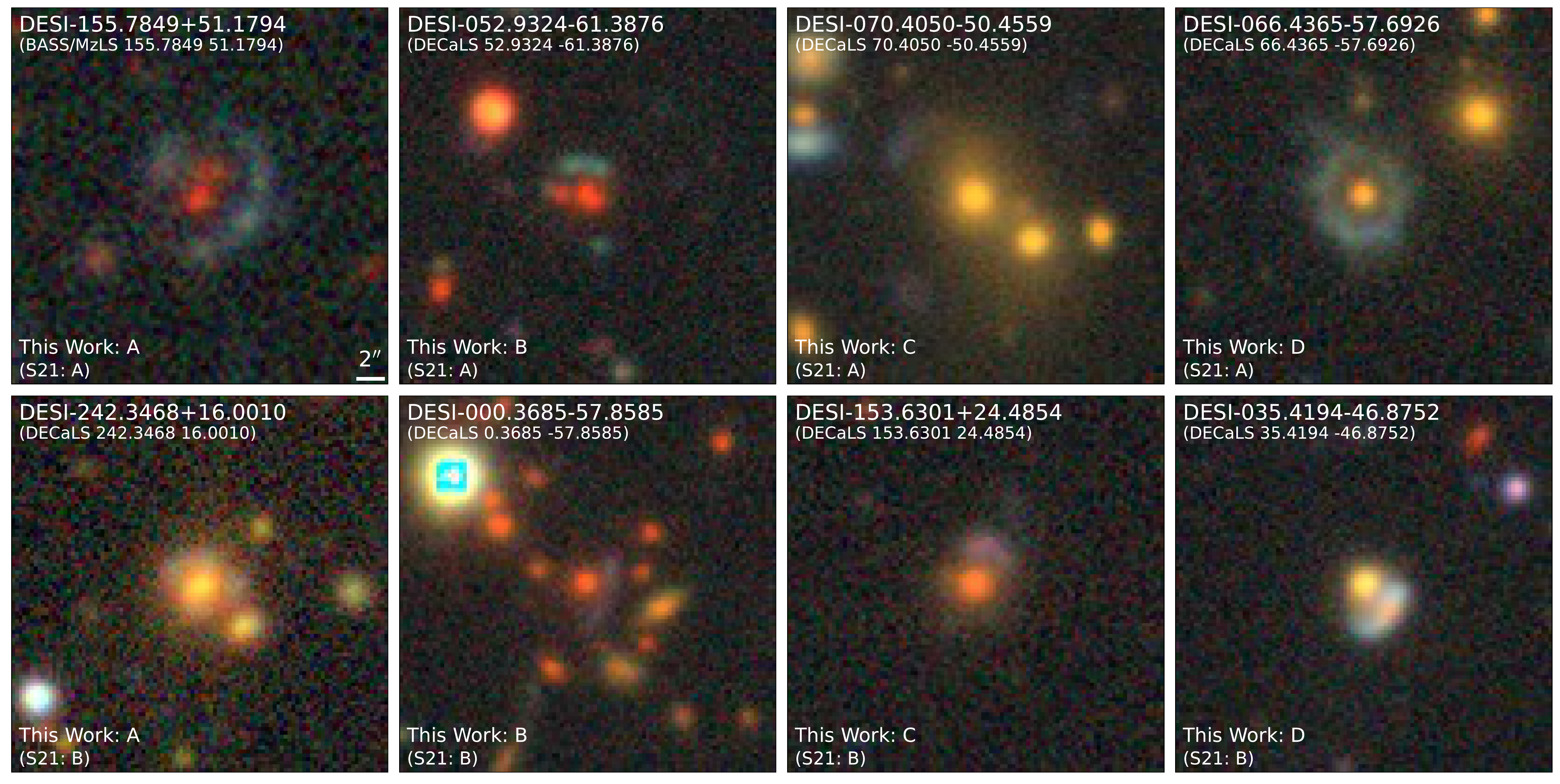}}
  % \captionsetup{font=footnotesize}
\captionof{figure}{\footnotesize A comparison of a sample of systems reported in both S21 and this work. \revs{For all images, north is up, and east to the left. The scalebar shown in the first image applies to all images in this figure.} We show an example system for all permutations of grade assignment given grades A and B from S21 and grades A, B, C, and D in this work. The name for each system from this work is shown in the top left, with the name from S21 \revs{in parentheses} below that. The bottom left shows the grade \revs{from this work and below that, the S21 grade in parentheses}.}
\label{fig:stein-comp}
\end{minipage}

\revs{Finally, the technique utilized by S21 is more computationally expensive (64 GPU hours, NVIDIA V100, with an additional 288 hours for their ``fine-tuned" approach) than our supervised approach (5 GPU hours, NVIDIA Tesla P100). On the other hand, the S21 approach has the advantage of facilitating searches for other kinds of objects (not just strong lensing systems) downstream.}
% In addition, they mention that computations were ``performed on 8 GPUs using Facebook AI Research’s Faiss11 library for efficient similarity search'', the amount of time that this step took \revs{is not stated in S21.}
% \revs{In addition, there was also 288 GPU hours used in their `fine-tuned" approach (S21, Section 4.2). }
% It is also unclear the contribution to the lens candidates from the ``fine-tuned" approach, which took an additional 288 GPU hours (S21, Section 4.2). 
% \revs{While the approach in S21 is novel and successful, it is burdened by much higher computational costs when compared with a typical self-supervised approaches.}
%``The pre-computed similarity array allows for rapid similarity searches of any image in the dataset; given a query image, the most similar images are immediately available to examine." -- did they say how long this ``pre-computation"  took?
\revs{Given future improvements on GPUs is almost a certainty, this is clearly is a promising strategy.} 

\revs{The discoverability of new lenses, by humans alone or with machine learning assistance, of course depends on observation conditions (e.g., depth, seeing, and filter selection) and the criteria for the deployment sample. As highlighted by the brief comparison between S21 and this work, even for the same survey, different well-designed machine learning based searches can lead to different discovery sets. In general, this likely is dependent on 
% for machine learning based searches, controlling factors include
the machine learning technique employed (e.g., neural network architecture, supervised vs. unsupervised), training sample (e.g., selection criteria, size, and whether it consists of simulations, observed images, or an amalgamation), and  the final human inspection process itself \citep{rojas2023}. These can lead to overlapping and complementary discoveries. A more detailed comparison between different searches and their respective results would likely be beneficial to future lens searches.} \revs{This, however,} is beyond the scope of this paper.

% A more comprehensive comparison between the approaches of S21 and ours and the respective results would likely be beneficial to future searches in large imaging surveys.

%%%%The overlap with candidates identified in each of the other recent publications \citep{canameras2021a, li2021a, rojas2021a, savary2021a,  odonnell2022a} are $\lesssim 0.5\%$ of our candidates.  
%%%%These has been noted in Table~\ref{tab:grade-A} and in our online tables.

% \newpage
\section{Conclusions}\label{sec:conclusion}
%%------------Conclusions----------------
We have carried out a search for strong gravitational lensing systems in the DESI Legacy Surveys Data Release 9 (DR9).
This is the third paper in a series on lens searches in the DESI Legacy Surveys, following Paper~I \citep[][DR7 DECaLS, {\tt DEV} and {\tt COMP} only]{huang2020a} and Paper~II \citep[][DR8, {\tt DEV}, {\tt COMP}, and {\tt REX}]{huang2021a}.
\begin{table}
\caption{Comparison Between S21 and Our Search\label{tab:stein-comp}}

% \footnotesize	
\scriptsize	
\begin{adjustwidth}{1cm}{-1.1cm}
\begin{tabularx}{\linewidth}{ |c| *{6}{Y|} |c *{1}{|Y} |c| *{1}{|Y|} }
%\begin{tabularx}{\linewidth}{ |c| *{6}{Y|} |c *{1}{|Y|} c| *{1}{|Y|} }

%\begin{tabularx}{\linewidth}{ |c| *{6}{Y|} |c *{1}{|Y} |c *{1}{|Y} |c *{1}{|Y} |c| *{1}{|Y} |c| *{1}{|Y|} }
% \begin{tabularx}{\linewidth}{ |c| *{6}{Y|} |c *{1}{|Y|} c *{1}{|Y} c *{1}{Y|} c| *{1}{|Y|}}

%\begin{tabularx}{\linewidth}{ |c|X|X|X|X|X|X|X|c |Y| c|Y| }

    \cline{1-11}
% \backslashbox{\color{ red}{\scriptsize S21}}{\color{blue} {\scriptsize This Work}}   &
\multicolumn{1}{|c|}{\backslashbox{\stein{S21}}{\our{This Work}}}                    &   
\multicolumn{2}{c|}{\our{A (115)}}&   \multicolumn{2}{c|}{\our{B (526)}}&   \multicolumn{2}{c||}{\our{C (1254)}}&   \multicolumn{1}{c|}{\our{ABC (1895)}} &   \multicolumn{2}{c||}{\our{D (1865)}} & \multicolumn{1}{c|}{\our{ABC+D}}\\
    \cline{1-11}
\multirow{2}{*}{\stein{A (404)}}                &          &   \stein{10.9\%}     &  & \stein{14.9\%}    &  & \stein{8.9\%} & \stein{34.7\%} & &  \stein{8.7\%} & \stein{43.3\%}  \\
\cline{2-11}
                                 &   \our{38.3\%}       &   44       & \our{11.4\%} & 60  & \our{2.9\%} & 36 & 140  & \our{1.9\%} & 35  & 175\\
\cline{1-11}

\multirow{2}{*}{\stein{B (788)}}                &          &   \stein{0.4\%}     &  & \stein{4.7\%}    &  & \stein{8.9\%} & \stein{14.0\%} & &  \stein{14.3\%} & \stein{28.3\%}  \\
\cline{2-11}
                                 &   \our{2.6\%}       &   3       & \our{7.0\%} & 37  & \our{5.6\%} & 70 & 110  & \our{6.1\%} & 113  & 223\\
% \cline{1-11}
\hhline{|===========|}

%\multirow{1}{*}{\stein{AB (1192)}}               
%\cline{2-11}
\stein{AB (1192)}        &   \our{40.9\%}       &   47       & \our{18.4\%} & 97  & \our{8.5\%} & 106 & 250  & \our{8.0\%} & 148  & 398\\

    \cline{1-11}
\end{tabularx}
\end{adjustwidth}
\tablecomments{Comparison of overlapping candidates discovered in S21 and our work.
The number of candidates from S21 for each of the two grades are shown in red.
Also in red are the overlapping systems by grade as a percentage of the number for the respective grade in S21.
Conversely, the number of candidates from our search are shown in blue.
Also in blue are the overlapping systems by grade as a percentage of the number in the respective grade \textit{in our work}.
For example, 44 \revs{are assigned Grade~A by this work and S21,} 
which is 10.9\% of the total number of Grade~A's (\steinAs) in S21 \revs{and 38.3\% of the Grade~A's reported in this work}.}
% Note that 10.4\% of the S21 Grade~A systems are given a Grade~D during our visual inspection;
% but we have not included in this table for the 44 overlapping Grade~D systems as a percentage of the total number of D's from our work, 
%Note that in the same row, for the 42 overlapping systems in our Grade~D category,
%we have not shown the corresponding percentage of the total number of D's from our work, since we do not report Grade~D systems in this paper.
\vspace{-3mm}

\end{table}

We use a customized deep residual neural network \citep{lanusse2018a, huang2021a},
trained on observed lenses and non-lenses.
We apply our trained neural network to $\sim$~\deploysize non-{\tt PSF} ({\tt SER}, {\tt DEV}, {\tt REX}, {\tt EXP}) 
cutout images with at least three passes in each of the $grz$ bands and a $z$-band magnitude cut of $< 20.0$ for the galaxy at the center of each image.
%%%% We use only real observations for training.  The smallest training sample and yet found lenses with one of the highest human inspection efficiency.
We hold a high standard in grading these candidate systems.
We have found \lensA Grade~A, \lensB Grade~B, and \lensC Grade~C candidates, for a total of \lenstot.
Of these, \lensprevknown were previously known systems, and \lensrecent were \revs{also reported} in recent publications, the majority of which came from \citet{stein2021a}. 
We therefore report \lenstotnew \textit{new} lens candidates \revs{with a grade breakdown of \newlensA Grade~A, \newlensB Grade~B, and \newlensC Grade~C.}
% This nearly doubles the number of lens candidates found by our group.
Combining all three searches, we have found a grand total of \grandtotnew strong lens candidates. 
% \lensaboveCplus of our candidates have a human inspection score $\geq 2.5$,
% all of which are at least likely lensing systems.
Grade~D systems are not counted as candidates in this paper, but we have included them on our project website (URL provided in \S\ref{sec:all-cands}). 

Along with \citet{stein2021a}, this is the largest deployment to search for strong lenses in the most heterogeneous dataset. 
% In addition, \revs{the dataset that we deployed} on is the most heterogeneous.  
%By using a statistically representative training sample, we have succeeded in finding many lenses with high efficiency.  
\revs{Even more notably,} in all parts of the footprint of our search, multiple lens searches were performed before, by our group or other groups.  
By using an expanded, statistically representative training sample, we are able to find a large number of high quality \textit{new} lens candidates, with high purity.
We have presented a brief comparison with \citet{stein2021a} in \S\ref{sec:discussion}.
Their results and ours show that a detailed comparison will likely benefit future lens searches in large data sets.
%(e.g., LSST, Euclid, the Roman Space Telescope).

%As the lens search efficiency has dramatically improved, 
\revs{This work, together with other searches, has clearly demonstrated that machine learning approaches are highly effective in discovering large numbers of high-quality strong lensing candidates. 
% \revs{Different search methods} have also been shown to \revs{yield} overlapping and complementary discoveries. 
This will likely continue to be the case in future surveys such as the Vera C. Rubin Observatory Legacy Survey of Space and Time (LSST), Euclid, and the Nancy Grace Roman Space Telescope, and the discovery of strong lenses will continue to accelerate.} There has also been significant development on strong lens modeling,
for example, the availability of open-source, widely-used lens modeling packages
%\lenstr 
\citep[e.g.,][]{birrer2018a, nightingale2019a}.
%is a fully-fledged (including, for example, a suite of plotting routines) and highly versatile lens modeling package.}
%However, the computational cost for lens modeling remains high.
Recently, \citet{gu2022a} introduced \gigalens, which takes advantage of the high level of parallelization of GPUs and \revs{automatic differentiation}, speeding up lens modeling by one to two orders of magnitudes.
This makes it possible to model the $\mathcal{O}(10^5)$ \revs{strong lensing systems expected to be discovered in the next decade} in a reasonable amount of time.
% when $\mathcal{O}(10^5)$ strong lensing systems are projected to be found in the next decade \citep{collett2015a}. and lens modeling greatly sped up \citep[e.g.,][]{gu2022a}, and larger and deeper surveys coming online, 
% Very recently, \citet{sheu2023a} conducted a retrospective search for lensed supernovae in the DESI Legacy Surveys data, and found seven promising candidates. 
The immediate future of using strong lensing to address significant astrophysical and cosmological questions is indeed very bright.

\newpage
\section{Acknowledgement}\label{sec:acknowledgement}
%%%% We thank 
%%%% We are grateful to Joel Brownstein and Lexi Moustakas for granting us access to the Master Lens Database (\url{http://admin.masterlens.org/index.php}).
This research used resources of the National Energy Research Scientific Computing Center (NERSC), a U.S. Department of Energy Office of Science User Facility operated under Contract No. DE-AC02-05CH11231 and the Computational HEP program in The Department of Energy's Science Office of High Energy Physics provided resources through the ``Cosmology Data Repository" project (Grant \#KA2401022).
%%This work was supported in part by the Director, Office of Science, Office of High Energy Physics of the U.S. Department of Energy under Contract No. DE-AC02-05CH11231.  
X.H. acknowledges the University of San Francisco Faculty Development Fund. 
A.D.'s research is supported by National Science Foundation's National Optical-Infrared Astronomy Research Laboratory, 
which is operated by the Association of Universities for Research in Astronomy (AURA) under cooperative agreement with the National Science Foundation.

This paper is based on observations at Cerro Tololo Inter-American Observatory, National Optical
Astronomy Observatory (NOAO Prop. ID: 2014B-0404; co-PIs: D. J. Schlegel and A. Dey), which is operated by the Association of
Universities for Research in Astronomy (AURA) under a cooperative agreement with the
National Science Foundation.

This project used data obtained with the Dark Energy Camera (DECam),
which was constructed by the Dark Energy Survey (DES) collaboration.
Funding for the DES Projects has been provided by 
the U.S. Department of Energy, 
the U.S. National Science Foundation, 
the Ministry of Science and Education of Spain, 
the Science and Technology Facilities Council of the United Kingdom, 
the Higher Education Funding Council for England, 
the National Center for Supercomputing Applications at the University of Illinois at Urbana-Champaign, 
the Kavli Institute of Cosmological Physics at the University of Chicago, 
the Center for Cosmology and Astro-Particle Physics at the Ohio State University, 
the Mitchell Institute for Fundamental Physics and Astronomy at Texas A\&M University, 
Financiadora de Estudos e Projetos, Funda{\c c}{\~a}o Carlos Chagas Filho de Amparo {\`a} Pesquisa do Estado do Rio de Janeiro, 
Conselho Nacional de Desenvolvimento Cient{\'i}fico e Tecnol{\'o}gico and the Minist{\'e}rio da Ci{\^e}ncia, Tecnologia e Inovac{\~a}o, 
the Deutsche Forschungsgemeinschaft, 
and the Collaborating Institutions in the Dark Energy Survey.
The Collaborating Institutions are 
Argonne National Laboratory, 
the University of California at Santa Cruz, 
the University of Cambridge, 
Centro de Investigaciones En{\'e}rgeticas, Medioambientales y Tecnol{\'o}gicas-Madrid, 
the University of Chicago, 
University College London, 
the DES-Brazil Consortium, 
the University of Edinburgh, 
the Eidgen{\"o}ssische Technische Hoch\-schule (ETH) Z{\"u}rich, 
Fermi National Accelerator Laboratory, 
the University of Illinois at Urbana-Champaign, 
the Institut de Ci{\`e}ncies de l'Espai (IEEC/CSIC), 
the Institut de F{\'i}sica d'Altes Energies, 
Lawrence Berkeley National Laboratory, 
the Ludwig-Maximilians Universit{\"a}t M{\"u}nchen and the associated Excellence Cluster Universe, 
the University of Michigan, 
{the} National Optical Astronomy Observatory, 
the University of Nottingham, 
the Ohio State University, 
the OzDES Membership Consortium
the University of Pennsylvania, 
the University of Portsmouth, 
SLAC National Accelerator Laboratory, 
Stanford University, 
the University of Sussex, 
and Texas A\&M University.

%%https://www.nersc.gov/users/accounts/user-accounts/acknowledge-nersc/
%%http://legacysurvey.org/

% \bibliographystyle{aasjournal}
%\clearpage
\bibliography{dustarchive}

% %%%% -----------  Tables and Figures for Candidates ---------

\end{document}